# Suppressed superconductivity in ultrathin Mo$_2$N films due to pair-breaking at the interface

M. Kuzmiak,[1,2] M. Kopčík,[1,2] F. Košuth,[1,3] V. Vaňo,[4] P. Szabó,[1] V. Latyshev,[3] V. Komanický,[3] and P. Samuely[1,3]

[1]*Centre of Low Temperature Physics, Institute of Experimental Physics, Slovak Academy of Sciences, SK-04001 Košice, Slovakia*
[2]*Department of Physics, Faculty of Electrical Engineering and Informatics, Technical University of Košice, SK-04001 Košice, Slovakia*
[3]*Centre of Low Temperature Physics, Faculty of Science, P. J. Šafárik University, SK-04001 Košice, Slovakia*
[4]*Department of Applied Physics, Aalto University, FI-00076 Aalto, Finland*

**Abstract**
A strong disorder characterized by a small product of the Fermi vector $k_F$ and the electron mean free $l$ drives superconductors towards insulating state. Such disorder can be introduced by making the films very thin. Here, we present 3-nm Mo$_2$N film with $k_{F*}l \sim 2$ with a resistive superconducting transition temperature $T_c$ = 2 K heavily suppressed in comparison with the bulk $T_c$. Superconducting density of states (DOS) with smeared gap-like peaks and in-gap states, so called Dynes DOS, is observed by the low temperature tunneling spectroscopy despite a sharp resistive transition. By scanning tunneling microscope the spectral maps are obtained and related to the surface topography. The maps show a spatial variation of the superconducting energy gap on the order of 20 % which is not accidental but well correlates with the surface corrugation: protrusions reveal larger gap, smaller spectral smearing and smaller in-gap states. In agreement with our previous measurements on ultrathin MoC films we suggest that the film-substrate interface introducing the local pair-breaking is responsible for the observed effects and generally for the suppression of the superconductivity in these ultrathin films.

*Magnetotransport, STM, tunneling spectroscopy, strongly disordered ultrathin superconducting Mo$_2$N films*

## 1 Introduction

Superconductivity is a macroscopic quantum phenomenon, where Cooper pairs forming the superconducting condensate can be described by a wave function $\Psi = \Delta\, e^{i\phi(r)}$, where $\Delta$ is the amplitude (related to energy gap $\Delta$) and $\phi$ is the phase. The increase of disorder to the critical value can destroy superconductivity in two different ways, either by suppressing the amplitude $\Delta$ (fermionic scenario [1] or by breaking the phase coherence $\phi$ (bosonic scenario [2]. In the case of fermionic scenario strong disorder leads to enhancement of Coulomb interaction eventually breaking the Cooper pairs into fermionic states with $\Delta \to 0$ at superconducting transition $T_c \to 0$, forming a bad metal. Then, at even higher disorder a second transition between bad metal and insulator can be accomplished. Thus, the superconductor-insulator transition (SIT) is a two-stage process through a bad metallic interstate. The bosonic scenario assumes a direct transition from the superconducting to the insulating state with eventually localized Cooper pairs. While the fermionic scenario predicts spatially homogeneous superconductivity before the transition, in the bosonic case due to the phase fluctuations the Cooper pairs spontaneously form „puddles" without global coherence and the superconducting gap value varies among these puddles. On the insulating side of bosonic SIT phase incoherent small superconducting islands float in the insulating sea [2, 3].

This paper presents experimental study of the superconducting properties of 3-nm thin Mo$_2$N film where a heavy disorder is introduced by a strongly reduced film thickness. The films with thicknesses smaller than 30 nm show gradually suppressed superconducting transition temperature on Mo$_2$N films in agreement with the finding in [4]. Our transport measurements showed that the SI transition takes place at thicknesses between 3 nm and 2 nm [5]. Thus, the 3-nm thin film is close to the critical disorder also characterized by a small product of the Fermi vector $k_F$ and the electron mean free path $l$ with $k_{F*}l \sim 2$. Superconducting density of states (DOS) directly measured by tunneling spectroscopy at very low temperatures yields smeared gap-like peaks and in-gap states, which are characteristic for the Dynes DOS [6]. Spectral maps obtained by scanning tunneling microscope (STM) show a spatial variation at the length scale of several nm with the superconducting energy gap changes on the order of 20 % which is not accidental but well correlates with the surface corrugation. The locally thinner parts of the film reveal smaller gap, larger spectral smearing and larger in-gap states, also the local $T_c$ is slightly decreased. In agreement with our previous measurements on ultrathin MoC films [7] we suggest that the film-substrate interface introducing the local pair-breaking is responsible for the observed effects and generally for the suppression of the superconductivity in these ultrathin films. It is an alternative to spontaneously emerging puddles which were reported on TiN ultrathin films where the bosonic superconductor-insulator scenario is assumed [3].

## 2 Experimental

Our Mo$_2$N films of a 3 nm thickness were prepared by reactive magnetron sputtering onto a sapphire substrate in the argon-nitrogen mixture. The subsequent crystallographic structure of the Mo$_2$N was characterized by XRD measurements. It was detected that our films form a stoichiometric γ-Mo$_2$N phase [8].

The critical temperature of the transition to the superconducting state has been determined from transport measurements. We measured the sheet resistance $R_s$ as a function of the temperature from 1.4 K to 300 K in an Oxford Instruments VTI system. Four electrical contacts have been prepared by painting a silver paint at the edges of the 5 x 5 mm$^2$ sample in the Van der Pauw configuration [9]. The value of the Ioffe-Regel parameter $k_F l \sim 2$ has been extracted from Hall effect and resistivity measurements and calculated from the free electron formula

$$k_F l = \frac{\hbar (3\pi^2)^{2/3}}{e^{5/3}} \left[ \frac{R_H^{1/3}}{\rho} \right], \quad (1)$$

where $\hbar$ is the reduced Planck constant, $e$ is the elementary electric charge, the resistivity $\rho = R_s t$ is the product of the sheet resistance $R_S$ and the thin film thickness $t$, $R_H$ is the Hall resistance and $n$ is the charge carrier density [10].

The surface and the local superconducting properties of the sample have been studied by the sub-Kelvin scanning tunneling microscope (STM) developed at the Center of Low Temperature Physics in Košice. This system enables STM experiments down to $T = 300$ mK and magnetic fields up to $B = 8$ T. A gold STM tip was used. First, the surface topography was measured and later at each point of the topography tunneling current-voltage (*I-V*) measurements were performed via the Current Imaging Tunneling Spectroscopy (CITS) technique [11]. The tunneling conductance $G(V) = dI(V)/dV$ has been calculated by a numerical differentiation of the *I-V* curves.

The tunneling conductance measured through the tunnel junction between a normal metal and an insulator can be described by the equation

$$G(H) = \frac{dI}{dV} = N_N \int_{-\infty}^{\infty} N_S \left[ \frac{-\partial f(E+eV)}{\partial eV} \right] dE, \quad (2)$$

where $I$ is the tunneling current, $E$ the energy, the derivative of the Fermi distribution in the square bracket represents thermal smearing. $N_N$ and $N_S$ are the densities of state (DOS) of the gold tip and the superconductor, respectively. The BCS theory defines $N_S$ as

$$N_S(E) = \mathrm{Re}\left(\frac{E}{\sqrt{E^2 - \Delta^2}}\right), \quad (3)$$

where $\Delta$ is the superconducting energy gap. Tunneling spectra with spectral smearing higher than thermal broadening can be described by the phenomenological modification of the BCS DOS of Dynes [6], who introduced to $N_S$ a complex energy $E = E' - i\Gamma$, where $\Gamma$ is the spectral smearing parameter of unknown origin.

The Dynes formula has been used to fit the tunneling spectra measured in our experiments. The gap-maps and gamma-maps have been constructed from the local $\Delta$ and $\Gamma$ values obtained from fitting of the tunneling spectra measured in the grid of 128x128 topography points.

## 3 Results

Figure 1 shows the basic characterization of the studied sample. Temperature dependence of the sheet resistance $R_s(T)$ is plotted in Fig. 1a). The main panel shows the $R_s(T)$ dependence from room temperature down to $T = 1.4$ K. About 10% increase of $R_s$ from the room temperature down to the superconducting transition is a typical signature of quantum corrections to Drude conductivity in a strongly disordered superconductor [12, 13]. The inset plots a detail of the superconducting transition. The critical temperature determined at 50% of the normal state $R^N_s = 900$ Ω is $T_c = 2$ K. The superconducting transition is very sharp with $\Delta T_c / T_c \approx 0.17$ ($\Delta T_c$ is the difference in $T_c$ determined at 90% and 10% of $R^N_s$) which confirms the high quality of the studied sample. Above the sharp transition $R_s$ slowly increases up to 6 K suggesting the presence of superconducting fluctuations typical for two-dimensional disordered systems and then a negative temperature derivative $dR_s/dT$ sets in.

A typical example of surface topography measured on a 300 x 300 nm$^2$ surface area at $T = 450$ mK is displayed in Fig. 1b). The surface shows a polycrystalline structure, which consists of 30 – 50 nm$^2$ large grains with average roughness of 0.1 nm. Maximum distance between valleys and protruded heights is about 1 nm as indicated in Fig. 1c).

Figure 2a) and b) show 3D color plots of typical temperature dependence of the tunneling spectra $dI/dV(V)$ at two different viewing angles. The spectrum measured at the lowest temperature $T = 500$ mK shows reduced coherence peaks and evident in-gap states. The smearing of the gap structure is much higher than the thermal broadening. During a very slow temperature drift a lot of spectra were taken very fast. At increasing temperature the superconducting gap-like structure is getting filled, i.e. the in-gap states are increasing, the gap-like peaks are diminishing and the gap is closing above 2.5 K as displayed in Fig. 2b). The value of superconducting energy gap $\Delta$ has been determined at each experimental temperature from fitting the tunneling spectra to the thermally smeared Dynes DOS. The fitting curve at the lowest temperature is plotted by the black solid line in Fig. 2a). The fit was obtained by variation of both the energy gap $\Delta$ and the spectral smearing $\Gamma$. At higher temperatures we could keep the low temperature value of $\Gamma = 0.13$ meV roughly constant with less than 10 % variation. The temperature dependence of $\Delta$ is shown by blue asterisks in Fig. 2c). Scattering of the data points is due to electronic noise and indicates the experimental error. The red line is a fit by the least square method to the BCS temperature dependence of the gap. The data satisfactorily follow the BCS prescription with $\Delta(0) = 0.44$ meV, and $T_c = 2.85$ K.

In the following, we focused our research on the local STM studies of superconducting properties of our sample. A typical surface topography measured on a smaller 20 x 40 nm$^2$ area is displayed in Fig. 3a). The color contrast highlights differences between deeper (blue) and higher (red) areas of the surface which are less than 1 nm. The protruding islands and valleys are about 5 to 7 nm wide. Three rows of islands are loosely stretching diagonally from left to right. On this surface the tunneling spectra have been collected in every point.

The evolution of the tunneling spectra measured along the black line marked in Fig. 3a) is plotted in Fig. 3b). The black line was chosen on the surface with relatively small corrugation. The figure presents a top view on the spectrum with its intensity given by the color. As can be directly seen even along the line with a small corrugation the in-gap states given by blue color are not really spatially homogeneous and the same holds for the gap-like peaks given by deep red color.

Evident spectral inhomogeneities are visible in the zero-bias-conductance (ZBC) map shown in Fig. 3c). Remarkably, one can discover a relation between the features in the ZBC-map and the topography map. Maybe the best (anti)correlation can be seen between two deep blue stains in ZBC-map showing the lowest ZBC (one in the bottom at about 16 nm and the second in the middle of the map) and two red protruding islands at the same places in Fig. 3a).

In the following we determined the local values of the superconducting energy gap $\Delta$ and the spectral smearing parameter $\Gamma$ from fitting all the tunneling spectra measured in the topography points of Fig. 3a) to the Dynes DOS. The resulting gap-map is shown in Fig. 3d) and gamma-map in Fig. 3e). The use of the same relative color contrast in these two maps indicates that the gap-map appears as a blurred negative of the gamma-map.

Comparing the surface topography (Fig. 3a)) with the gap-map (Fig. 3d)) we can see, that in 20-40 nm$^2$ large red areas protruding above the deepest parts of the surface the value of the superconducting energy gap $\Delta$ is 20% higher ($\Delta \approx 0.46$ meV), than in the lower blue areas ($\Delta \approx 0.37$ meV). Concurrently, the $\Gamma$ values show the opposite tendency. $\Gamma$ is lower at the top of the protruding surface areas ($\Gamma \approx 0.14$ meV) compared to the deepest parts ($\Gamma \approx 0.26$ meV).

More quantitatively the relation between the local gaps $\Delta$ and the related smearing parameter $\Gamma$ is presented in Fig. 3f), where the $\Delta-\Gamma$ density plot is constructed from the data shown in Fig. 3d) and Fig. 3e). The values of the energy gap are decreasing linearly with increasing the spectral smearing providing evidence for their anticorrelation.

We can summarize from the presented experiments that spatial variations of the superconducting gap, smearing parameter and the zero-bias tunneling conductance are not accidental but are closely related to the surface corrugations. The thicker is locally the sample, the larger is the superconducting energy gap $\Delta$, the smaller spectral smearing $\Gamma$ and the smaller the zero-bias conductance.

## 4 Discussion

The influence of film thickness on the superconducting properties of homogeneously disordered ultrathin MoC films was studied in our previous papers. Low temperature STM measurements [14] showed that lowering the film thickness from 30 nm to 3 nm leads to a suppression of superconducting transition temperature. Reduced $T_c$ upon increased disorder with decreased thickness of the film is supposed to be a result of a renormalization of the Coulomb interaction. In accordance with the widely used Finkelstein model [1] based on the renormalization group analysis a decreasing $T_c$ is directly driven by the increase of the sheet resistance $R_s$. As shown in Ref. [14] dependence of the superconducting transition on the sheet resistance $T_c(R_s)$ of a series of MoC films with different thickness could be well fitted by the Finkelstein model. But this model was not able to explain two other findings: first, the introduction of the finite value of the spectral smearing parameter $\Gamma$ in the samples thinner than 10 nm and second, the fact that two 3-nm thin MoC samples prepared simultaneously on two different substrates featuring the same sheet resistance revealed different superconducting parameters. In particular, in the paper [7] we showed that 3-nm MoC film deposited on silicon substrate featured higher $\Delta$ but smaller $\Gamma$ than 3-nm MoC film prepared simultaneously on the sapphire and showing the same scalar disorder evidenced by the same sheet resistance.

Spectral broadening in the density of states is evidenced in many superconductors. As mentioned already in the original paper of Dynes [6] it can be a consequence of anisotropy effects, noise, or concentration fluctuations. In some case it can have an extrinsic origin. It can reflect the sample inhomogeneity and then be case-dependent and its relative size of $\Gamma/\Delta$ can vary from place to place when measured at different tunneling junctions [15]. A generic description of the origin of the $\Gamma$ in Dynes formula has been recently published by Herman and Hlubina [16]. In their microscopic interpretation the smearing $\Gamma$ is connected with the Lorentzian distribution of local pair-breaking fields in arbitrary potential disorder.

What can be a source of such pair-breaking fields in our ultrathin films? Could it be a tiny contamination by magnetic impurities for example? Remarkably, it was shown that finite smearing was not observed on the 30 nm

MoC film but appeared only on thinner samples with thickness comparable or smaller than the coherence length of the film despite the fact that the starting materials and growth procedure were the same. This excludes a simple contamination. On the series of 30, 10, 5 and 3 nm MoC samples it was shown that the thinner the sample, the smaller is $T_c$ and $\Delta$ and bigger $\Gamma$. Even more, on the single 3-nm MoC sample well defined protrusions exceeding the rest of the sample by about 1 nm also showed higher $\Delta$ and smaller $\Gamma$.

Experimental data of the 3-nm $Mo_2N$ film presented in this paper are very similar to the above-mentioned results obtained on MoC ultrathin films. The surface topography (Fig. 3a)) indicates correlations with the local variation of the superconducting energy gap in Fig. 3d) and anticorrelations with the spectral smearing parameter (Fig. 3e)). In thinner areas $\Delta$ is smaller and $\Gamma$ higher than in the surface of the protruding areas. It all strongly suggests that the pair-breaking fields reside at the interface between the sapphire substrate and $Mo_2N$ film. Figure 3f) displays in the density plot a linear suppression of the energy gap $\Delta$ with increased spectral smearing $\Gamma$. Several STM measurements of the temperature dependence of the energy gap showed that it is not only the gap value which is changing but also local $T_c$ follows it, keeping the coupling strength $2\Delta/k_BT_c$ constant and close to the BCS value of 3.52. Then, the pair-breaking reflected by $\Gamma$ is indeed an important factor in suppression of superconductivity, here and will be acting together with enhanced Coulomb interaction.

Possible sources of the local magnetic pair breaking can be the interface areas, where the diffusion of oxygen atoms form uncompensated electron orbitals. Gapless tunneling DOS have been detected on Nb where substoichiometric $Nb_2O_5$ oxides on the surface produced unscreened $d$-band magnetic moments [17]. While diamagnetic in bulk many nanomaterials and thin films reveal a room temperature ferromagnetism, incl. sapphire [18]. It was shown that ferromagnetism observed in $Al_2O_3$ samples arises from ferromagnetic coupling of Al vacancies [19]. Very recently, Tamir et al. [20] have reported that magnetic disorder is also present at the surface of amorphous superconducting films without any magnetic impurities but related to the surface termination. Many disordered superconducting films exhibit smeared tunneling characteristics or more generally, show strong dissipation deep in the superconducting state. We can resume that the local magnetic moments at the interface between the substrate and films can be a source of such dissipation.

## 5 Conclusions

The superconducting properties of strongly disordered 3-nm thin $Mo_2N$ films have been studied by transport and low temperature STM measurements. Transport show a very sharp transition to the superconducting state indicating homogeneous distribution of disorder but local STM measurements discover a spatial variation of the spectral characteristics which is related to the topography corrugation where the hills and valleys with a lateral size of about 5 to 7 nm differ up to 1 nm in height. The thinner the film, the lower the superconducting energy gap $\Delta$ and the larger the smearing parameter $\Gamma$. The energy gap $\Delta$ decreases linearly with the increasing spectral smearing $\Gamma$ evidencing that $\Gamma$ is important pair-breaking parameter responsible at least partially for suppression of superconductivity and accelerating superconductor-insulator transition. The most probable source of the pair-breaking fields is the interface between the sapphire substrate and the MoC film where uncompensated electron orbitals can produce local magnetic moments. Our results suggest that the spatial inhomogeneities of the superconducting characteristics are related to a spatial variation of the thickness, i.e. variation of the distance to pair-breaking fields at the interface. It is very different effect from a spatial variation of the superconducting gap caused by a spontaneous formation of superconducting puddles in the bosonic superconductor-insulator transition.

**Acknowledgments** We gratefully acknowledge helpful conversations with M. Grajcar and R. Hlubina.

**Funding Information** This work was supported by the projects APVV-18-0358, VEGA 2/0058/20, VEGA 1/0743/19 the European Microkelvin Platform, the COST action CA16218 (Nanocohybri) and by U.S. Steel Košice.

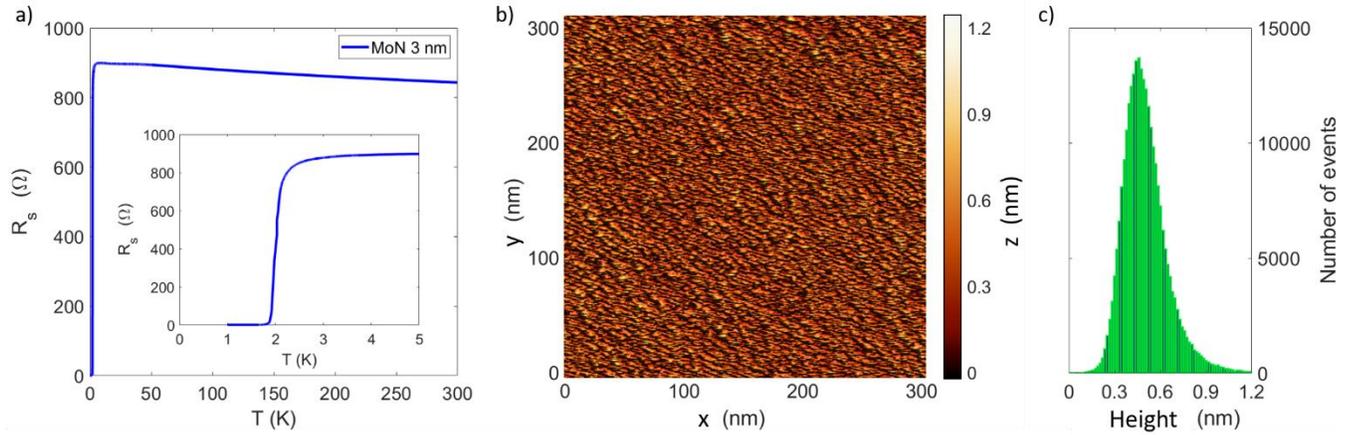

Fig. 1. **Basic characterization of 3-nm Mo₂N thin film.** a) Temperature dependence of the sheet resistance $R_s(T)$ with zoom of the transition around the transition in the inset. b) Surface topography of Mo₂N-3nm obtained at $V = 42$ mV, $I = 3.4$ nA, $T = 450$ mK. The image size measured by the STM is $300 \times 300$ nm². c) Histogram of heights of the surface from b).

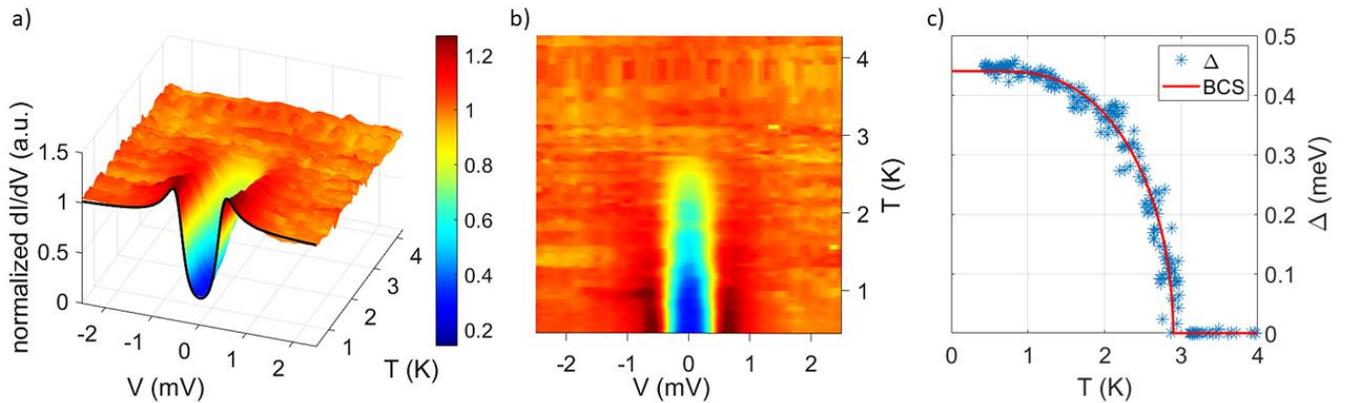

Fig. 2. **Tunneling spectroscopy.** a) The 3D plot shows a typical temperature dependence of the STM differential conductance spectrum $dI/dV$ of 3-nm Mo₂N film at measuring parameters $V = 3$ mV, $I = 0.4$ nA. All spectra were normalized to their values in the normal state at bias voltage $V = 3$ mV. The black line shows the fitting curve to Dynes model at $T = 500$ mK. b) shows a top view of the 3D plot from a). c) Temperature dependence of the energy gap $\Delta(T)$ obtained from fitting of all spectra from a) to the Dynes model. The red line shows predictions of the BCS theory with $\Delta(0) = 0.44$ meV and $T_c = 2.85$ K.

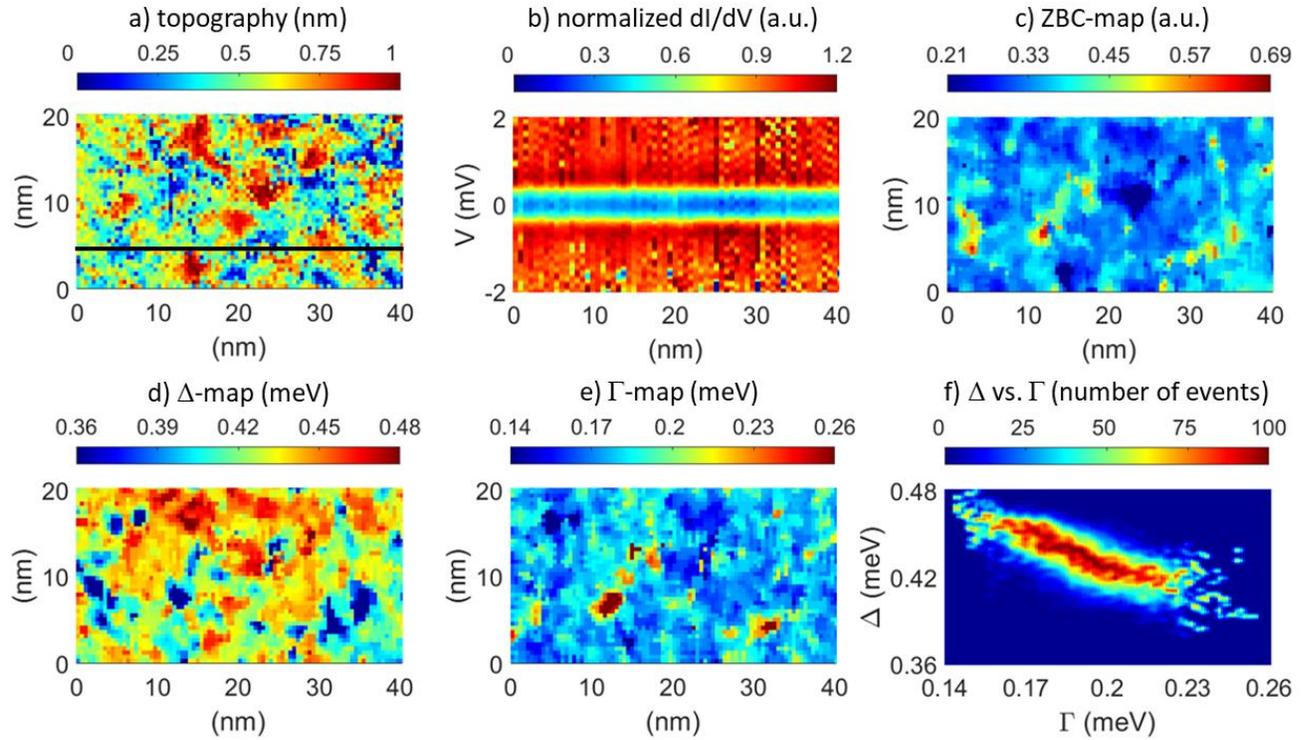

Fig. 3. **Current imaging tunneling spectroscopy (CITS) results on 3-nm Mo$_2$N film.** a) Surface topography measured by STM at parameters $V = 3$ mV, $I = 0.4$ nA, and $T = 500$ mK. b) Top view of the STM differential conductance spectra $dI/dV(V)$ measured along a 40 nm long line plotted in the topography image, c) Zero-bias-conductance map, d) $\Delta$–map, e) $\Gamma$-map, f) $\Delta$–$\Gamma$ density plot constructed from $\Delta$ and $\Gamma$ values shown in b) and c), respectively.